\documentclass{moriond}

\usepackage{graphicx}
\usepackage{placeins}
\usepackage{relsize}
\usepackage{lineno}

\def\PRL{\em \PRL}

\def\be{\begin{equation}}
\def\ee{\end{equation}}
\def\bea{\begin{eqnarray}}
\def\eea{\end{eqnarray}}

\newcommand{\Photo}{}

\def\babar{\mbox{\slshape B\kern-0.1em{\smaller A}\kern-0.1em B\kern-0.1em{\smaller A\kern-0.2em R}}}

\begin{document}
\vspace*{4cm}
\title{Tau and low multiplicity physics at Belle
and Belle II}

\author{ L. Corona~\footnote{on behalf of the Belle~II Collaboration}}

\address{INFN Sezione di Pisa, I-56127 Pisa, Italy}

\maketitle \abstracts{The Belle and Belle II experiments have collected a 1.4~ab$^{-1}$ sample of $e^+e^-$ collision data at centre-of-mass energies near the $\Upsilon (nS)$ resonances, of which 424~fb$^{-1}$ were collected at Belle~II in Run1 (2019--2022). 
We present a measurement of the lepton-flavour universality between electrons and muons, the search for the lepton-flavour violation decay $\tau^{-} \to \mu^{-}\mu^{+}\mu^{-}$, and a measurement of the $e^+e^- \to \pi^+\pi^-\pi^0$ cross section in the energy range from 0.62--3.5~GeV using an initial-state radiation technique, all performed with Belle~II data.} 

\section{Introduction}
Belle~II~\cite{B2TIP} is a high-intensity frontier experiment that operates at the SuperKEKB $e^+e^-$ asymmetric-energy collider~\cite{SKEKB}. The Belle~II detector~\cite{B2TDR} and the SuperKEKB collider are major upgrades of the Belle detector~\cite{BTDR} and the KEKB collider~\cite{KEKB}.  
The Belle and Belle~II experiments have collected a 1.4~ab$^{-1}$ sample of $e^+e^-$ collision data at centre-of-mass energies near the $\Upsilon (nS)$ resonances, of which 424~fb$^{-1}$ were collected at Belle~II in the first data taking run (Run1), from 2019 to 2022.
The second period of data taking (Run2) at Belle~II has started on February 20$^{\rm th}$ 2024.

The Belle~II data sample collected in Run1 contains 389 million $e^+e^- \to \tau^+\tau^-$ events.
In the centre-of-mass frame, $\tau$ pairs are produced back-to-back, with the decay products of each $\tau$ isolated from those of the other $\tau$ and contained in two opposite hemispheres. The boundary between the hemispheres is taken as the plane perpendicular to the thrust axis direction, $\hat{n}_{T}$, which is defined as the direction that maximizes the thrust value $\sum_i|\hat{n}_{T}\cdot \overrightarrow{p_i}^{\ast}|/\sum_i|\overrightarrow{p_i}^{\ast}|$, where $\overrightarrow{p_i}^{\ast}$ is the momentum of each final state particle expressed in the centre-of-mass frame. The decay products of one of the two $\tau$s are used to select the events (\textit{tag} side), and we search for signal in the decay products of the other $\tau$ (\textit{signal} side). 
To suppress background, we reconstruct specific topologies. For example, in the $1 \times 3$ topology, we reconstruct 1 charged particle in the \textit{signal} side and 3 charged particles in the \textit{tag} side. 
We use the collected $e^+e^- \to \tau^+\tau^-$ events for precision tests of the SM~\cite{taumass}, and the search for new physics in $\tau$ decays~\cite{taula}. 

Finally, thanks to the constrained kinematics, a low background environment, and dedicated triggers, Belle~II has excellent reconstruction capabilities for low multiplicity signatures~\cite{snow}.

\section{Lepton-flavour universality in \texorpdfstring{$\tau$}{} decays}
In the Standard Model (SM) of particle physics, the lepton-flavour universality (LFU) refers to the property of the electroweak gauge bosons to couple to the three generations of leptons through the same coupling. 
However, several extensions of the SM postulate the existence of new particles with different couplings to the three leptons.

We present a measurement of the $e-\mu$ lepton-flavour universality by comparing the measured branching ratios of the leptonic $\tau$ decays $\mathcal{B}(\tau^- \to \mu^-\bar{\nu}_{\mu}\nu_{\tau})$ and $\mathcal{B}(\tau^- \to e^-\bar{\nu}_{\mu}\nu_{\tau})$ through the ratio 
$R_{\mu} = \mathcal{B}(\tau^- \to \mu^-\bar{\nu}_{\mu}\nu_{\tau})/\mathcal{B}(\tau^- \to e^-\bar{\nu}_{\mu}\nu_{\tau})$.
The $R_{\mu}$ ratio constraints the ratio    
$\left\vert g_{\mu}/g_{e}\right\vert_{\tau} = \sqrt{R_{\mu}\cdot f(m^{2}_{e}/m^{2}_{\tau})/f(m^{2}_{\mu}/m^{2}_{\tau})}$,  where $g_{e}$ and $g_{\mu}$ are the $W^{\pm}$-$e$ and $W^{\pm}$-$\mu$ couplings, respectively; $m_{e}$, $m_{\mu}$ and $m_{\tau}$ are the masses of the corresponding leptons; and $f(x) = 1 - 8x + 8x^2 - x^4 -12x^2ln(x)$, assuming negligible neutrino masses~\cite{lfu}.

We use a data sample of 362~fb$^{-1}$ collected at the $\Upsilon(4S)$ resonance at Belle~II during Run1. We reconstruct the $1\times1$ topology with one electron or one muon in the \textit{signal} side, and one charged hadron and at least one $\pi^0$
in the \textit{tag} side. In this topology, we profit from the high branching fractions of $\tau^- \to h^-n\pi^0\nu_{\tau}$, 
the low backgrounds, and the high trigger efficiency of 96.6\% and 99.8\% for the $\tau^- \to \mu^-\bar{\nu}_{\mu}\nu_{\tau}$ and $\tau^- \to e^-\bar{\nu}_{\mu}\nu_{\tau}$ decays, respectively.

The main backgrounds are QED radiative di-lepton and four-lepton final-state processes and two-photon processes with more than two hadrons in the final state. The two-photon processes are suppressed by applying a data-driven selection based on thrust value, missing mass and missing momentum. The other backgrounds are suppressed using a neural-network fed with the magnitude and polar angle of the thrust vector, the visible energy in the centre-of-mass frame, the transverse component of the missing momentum direction in the centre-of-mass frame, and kinematic variables sensitive to the $h^-n\pi^0$ system in the \textit{tag} side. The main remaining backgrounds are the following: 3.3\% of $e^+e^- \to \tau^+\tau^-$ with one of the $\tau$s decaying to a $\pi$ and the $\pi$ faking a muon or an electron in the \textit{signal} side; 
2.5\% of $e^+e^- \to \tau^+\tau^-$ events with a mis-identified \textit{tag};  
and 0.2\% of $e^+e^- \to \tau^+\tau^-\tau^+\tau^-$. 
The signal reconstruction efficiency is 9.6\% with 94\% purity for the combined $e-\mu$ sample.

The $R_{\mu}$ ratio is measured through template binned maximum likelihood fits over 21 lepton momentum bins from 1.5 to 5~GeV$/c$.  The templates for the signal and background 
yields are derived from simulation. The main systematic uncertainties are 0.36\% from particle identification and 0.10\% from trigger efficiency. 
The total systematic uncertainty is 0.37\%, which is included in the fit as nuisance parameter. We check the stability of the results selecting different data periods, applying different selections on particle identification variables, electron and muon kinematic variables, and neural network output,  
and changing the number of bins used to define the templates. We obtain consistent results in all cases,  
indicating no significant unaccounted-for systematic effects.  

\begin{figure}[!ht] 
  \centering
    \includegraphics[width=0.45\linewidth]{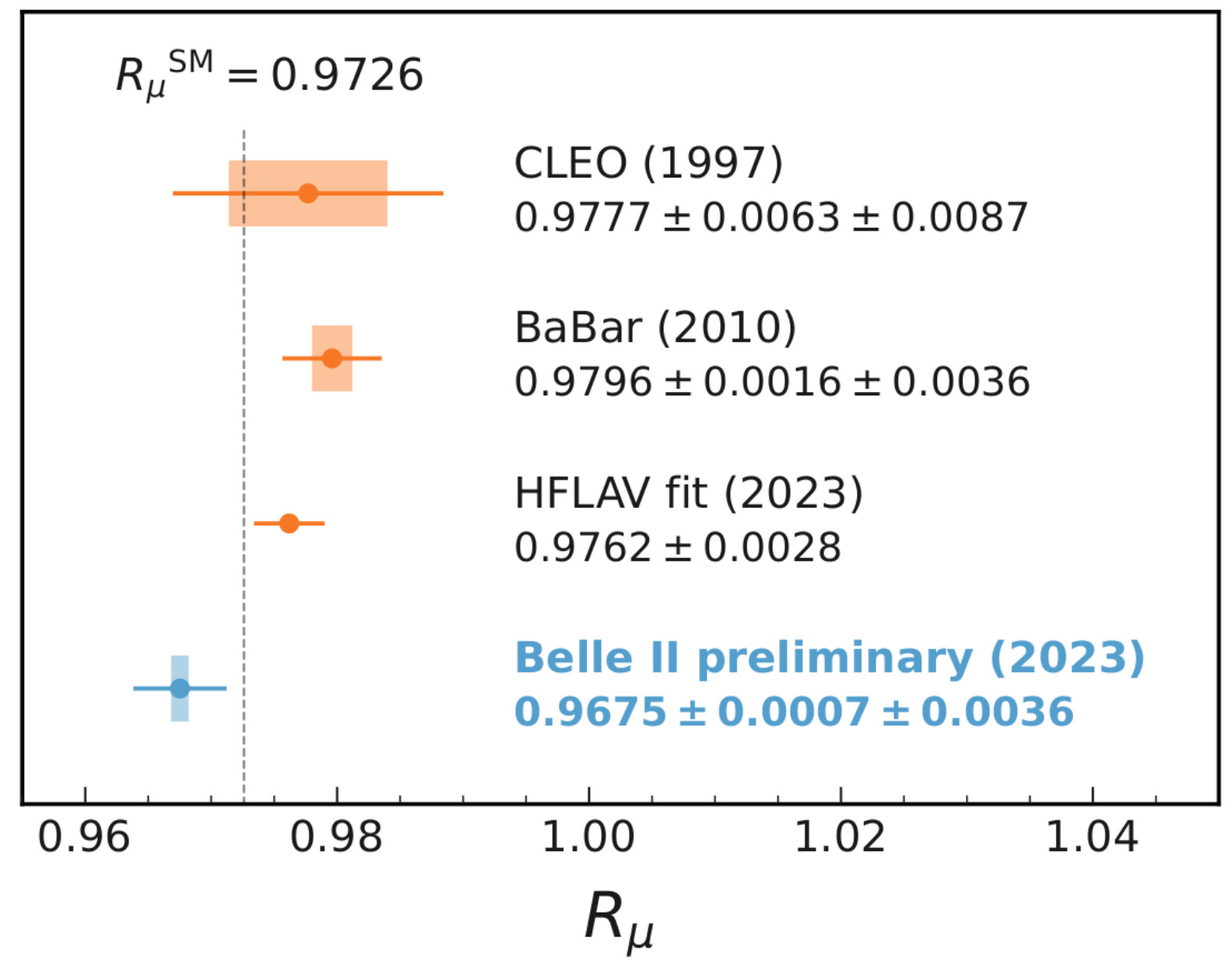}
    \includegraphics[width=0.45\linewidth]{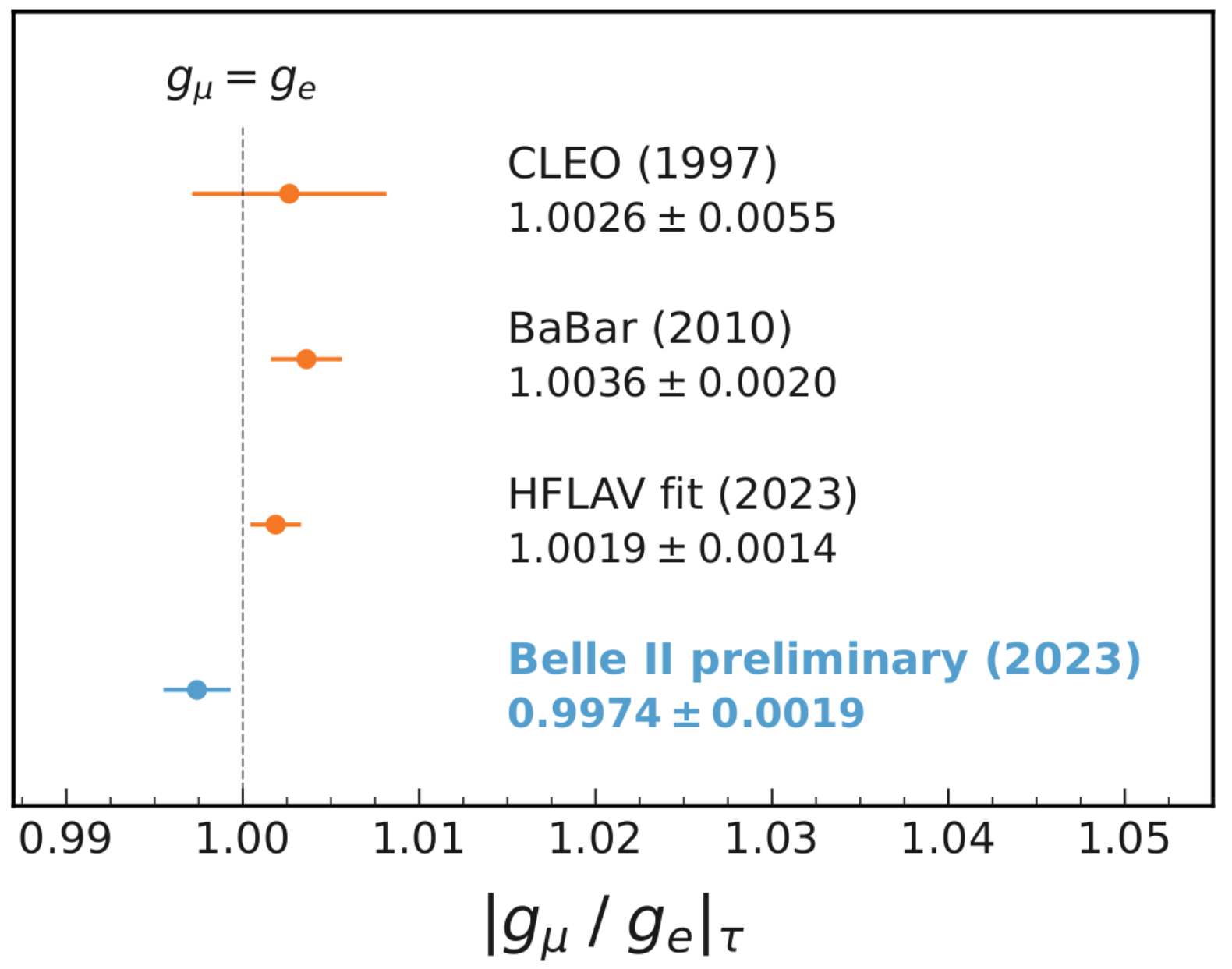} 
    \caption{Results obtained for $R_{\mu}$ (left) and $\left\vert g_{\mu}/g_{e} \right\vert_{\tau}$ (right) from Belle~II measurement (light blue) compared with previous measurements~\protect\cite{lfucleo,lfubabar}, and the fit from the Heavy Flavor Averaging Group~\protect\cite{hflav} (orange). The shaded areas in the left plot represent the statistical uncertainties, while the error bars indicate the total uncertainties. The vertical dashed line indicates the SM prediction, including mass effects.}
    \label{fig:lfu}
\end{figure}

We measure $R_{\mu} = 0.9675 \pm 0.0007(stat.) \pm 0.0036(syst.)$, which is converted in the most stringent test of LFU in $\tau$-lepton decays from a single experiment obtaining $\left\vert g_{\mu}/g_{e}\right\vert_{\tau} = 0.9974 \pm 0.0019$. Results are shown in Figure~\ref{fig:lfu}. From the combination of CLEO, \babar\ and Belle~II yields, assuming independent systematic uncertainties, we obtain $\left\vert g_{\mu}/g_{e}\right\vert_{\tau} = 1.0005 \pm 0.0013$.

\section{Search for the lepton-flavour violation decay \texorpdfstring{$\tau^- \to \mu^-\mu^+\mu^-$}{}}
The charged-lepton-flavour violation (LFV) is predicted in many extension of the SM, but it has never been observed. The conservation of the flavour of charged leptons is an accidental symmetry in the SM, and it is only broken at loop level when taking into account neutrino mixing. In the SM extended with massive neutrinos, the LFV effects are proportional to the very small neutrino masses, with decays  
suppressed at the order of $10^{-50}$. 
However, LFV effects are enhanced in several extensions of the SM at the level of $10^{-10}$--$10^{-8}$, thus an observation would be a clear signature of new physics.

We present the search for the charged LFV violation decay $\tau^- \to \mu^-\mu^+\mu^-$ using the full Run1 data sample collected at Belle~II, corresponding to an integrated luminosity of 424~fb$^{-1}$. This specific channel is experimentally favored because it is neutrinoless, so the $\tau$ mass and energy can be precisely determined, and the SM background sources are suppressed. The best upper limits on this search are from Belle~\cite{lfvbelle}, which performed the analysis using 782~fb$^{-1}$ and set upper limits of $2.1\times10^{-8}$ at 90\% confidence level. 
Belle has reconstructed a $3\times1$ topology of $e^+e^- \to \tau^+\tau^-$ events, with three muons in the \textit{signal} side and one charged particle 
in the \textit{tag} side. Despite the smaller data sample, Belle~II is already competitive  
by applying an inclusive approach on the \textit{tag} side, i.e. without reconstructing a specific $\tau$ decay. 

We search for signal 
in the ($M_{3\mu},\Delta E_{3\mu}$) plane, where $\Delta E_{3\mu} = \sqrt{s}/2 - E_{\tau,sig}$, $E_{\tau,sig}$ is the reconstructed energy of the \textit{signal} $\tau$,  $\sqrt{s}$ is the energy in the $e^+e^-$ centre-of-mass frame, and $M_{3\mu}$ is the invariant mass of the three muons. Since $\tau^- \to \mu^-\mu^+\mu^-$ decay is neutrinoless, $\Delta E_{3\mu}$ and $M_{3\mu}$ are expected to peak at zero and at the nominal mass of the $\tau$, respectively. 
The final signal yield is extracted from a $\pm5\delta$ semi-axis wide asymmetric elliptical region centered at the signal peak in the ($M_{3\mu},\Delta E_{3\mu}$) plane, where $\delta$ is the width obtained from a fit to both $M_{3\mu}$ and $\Delta E_{3\mu}$ variables. The sideband region is defined as a rectangular area covering the region ($\pm20\delta_{M},\pm10\delta_{\Delta E}$) excluding the signal region.

Several background processes can be mis-identified as signal candidates: QED radiative di-lepton and four-lepton final-state processes, $e^+e^- \to q\bar{q}$, $e^+e^- \to e^+e^-h^+h^-$ with $h = \pi, K, p$, and two photon processes with more than two hadrons in the final state.  
The signal efficiency and background suppression are optimized by using boosted decision trees (BDT) fed with kinematic variables related to the \textit{signal} $\tau$, variables related to the properties of all the particles that are not used in the signal reconstruction, and event shape variables. The selection on the BDT output is optimized by maximizing a Punzi figure-of-merit~\cite{punzi}. The signal efficiency is $20.43\% \pm 0.06\%(stat.)$ for a background reduction of 98.2\%.
The number of expected background events in the signal region is estimated from data in the sideband region, and it turns out to be $0.50^{+1.4}_{-0.5}$, where uncertainties are statistical.
The largest systematic uncertainty is 16\%, and it is due to imperfections in the magnetic field description and material mismodeling, which affects the number of expected background estimated from the sideband region.

We validate the inclusive reconstruction approach and the BDT selection using the $3\times1$ approach, i.e. reconstructing one charged particle in the \textit{tag} side.  
We obtain that the inclusive tagging selection has a 37\% higher efficiency than the $3\times1$ topology approach for a similar level of background expected, so the inclusive approach is preferred for the final result.

\begin{figure}[!ht] 
  \centering
    \includegraphics[width=0.45\linewidth]{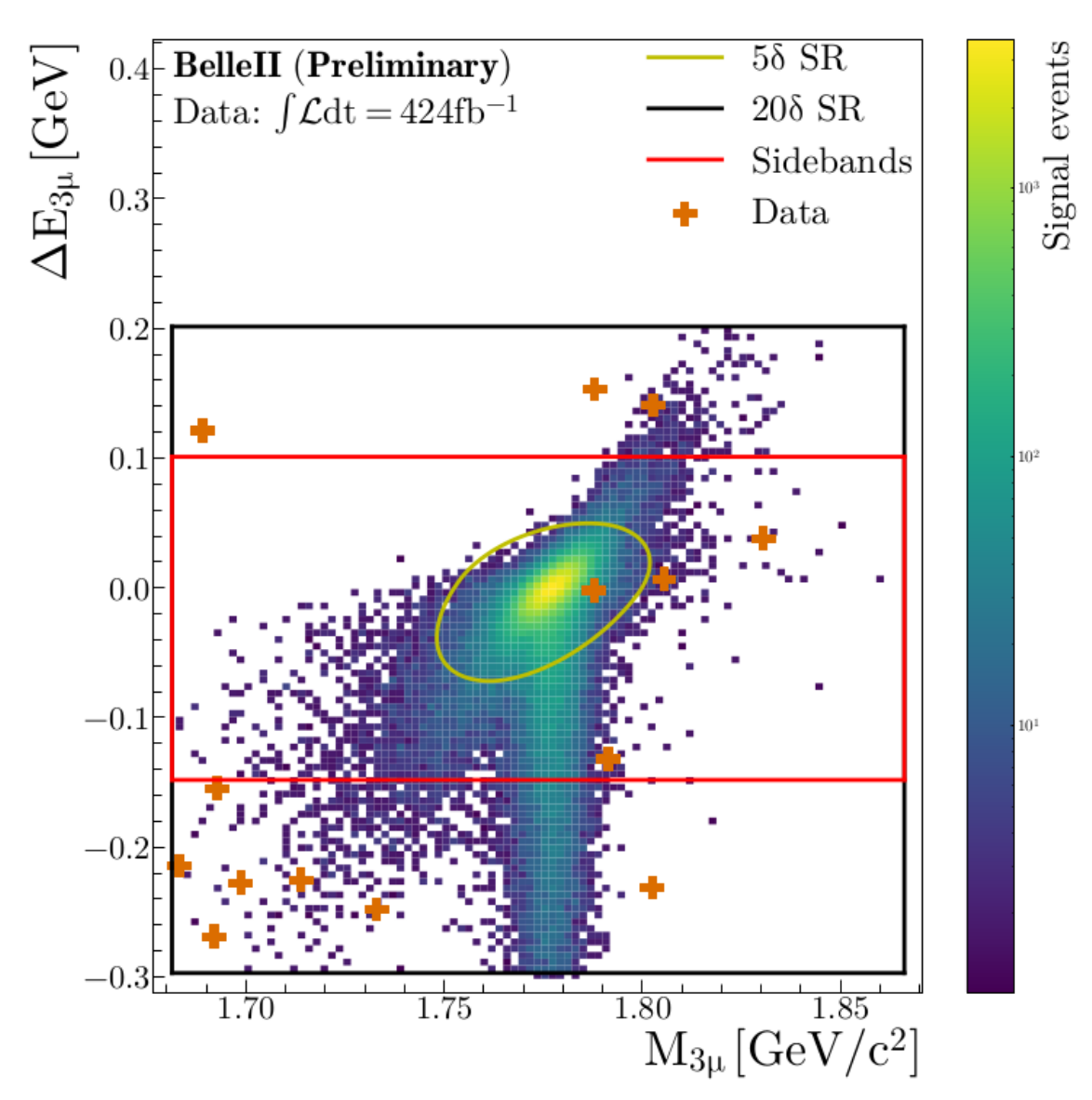}
    \includegraphics[width=0.54\linewidth]{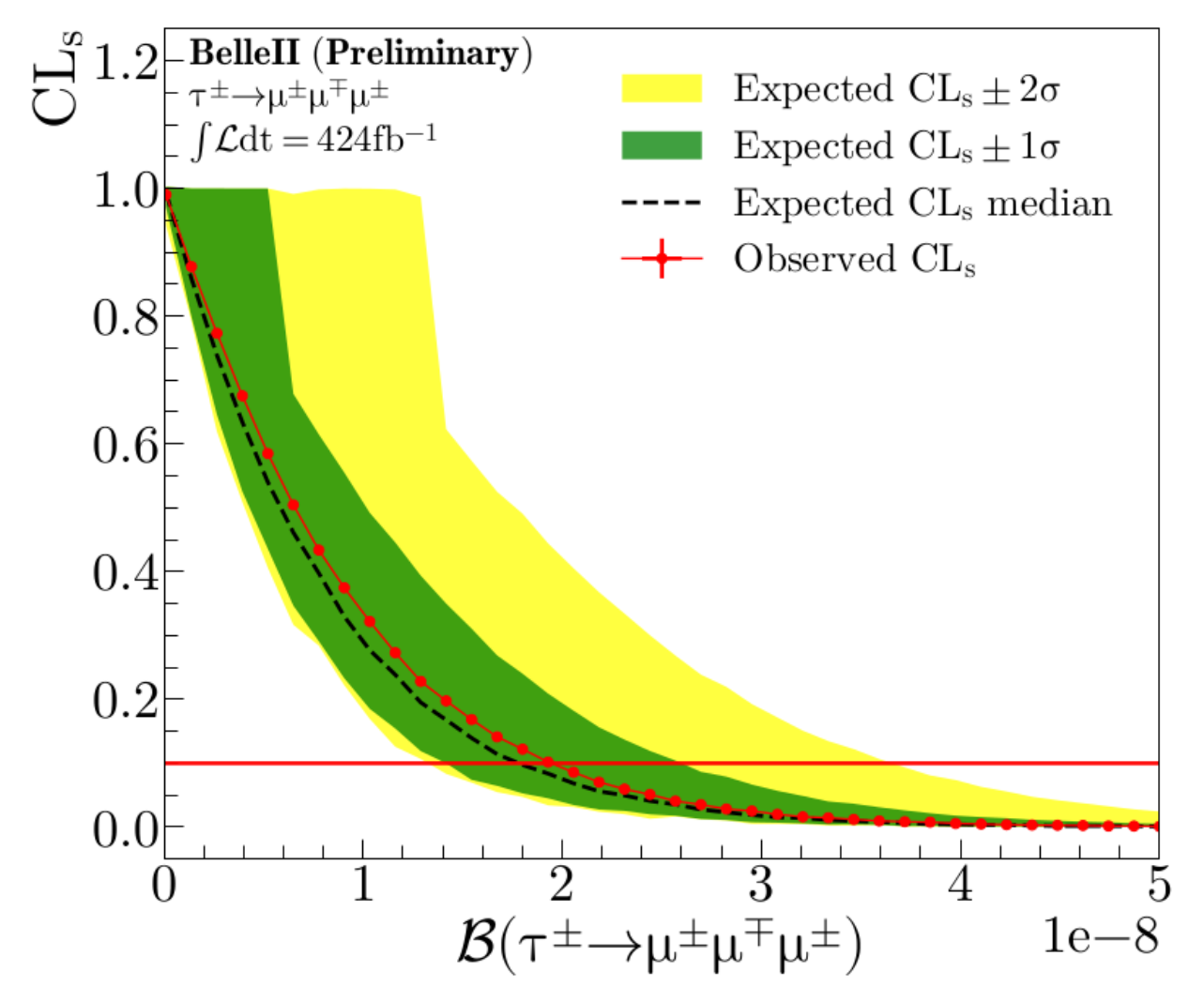} 
    \caption{Left: distribution of observed events in data (orange points) and simulated signal events (color filled area) in the ($M_{3\mu},\Delta E_{3\mu}$) plane. Sideband and signal regions are shown as a red rectangle and a yellow ellipse, respectively. Right: observed and expected $CL_S$ curves as a function of the upper limit on $\mathcal{B}(\tau^-\to\mu^-\mu^+\mu^-)$. The red line corresponds to the 90\% confidence level.}
    \label{fig:lfv}
\end{figure}

In data, we observe one event in the signal region, which is compatible with the background expectation. The corresponding measured branching ratio is $\mathcal{B}(\tau^-\to\mu^-\mu^+\mu^-) = (3.1^{+8.7}_{-3.6}(stat.)\pm0.1(syst.))\times10^{-9}$.
We do not observe any significant excess and set the most stringent limits at 90\% confidence level of $1.9\times10^{-8}$ on $\mathcal{B}(\tau^-\to\mu^-\mu^+\mu^-)$.
Figure~\ref{fig:lfv} shows the distribution of observed events in data and the simulated signal events in the ($M_{3\mu},\Delta E_{3\mu}$) plane (left), and the observed and expected $CL_S$ curves as a function of the upper limit on $\mathcal{B}(\tau^-\to\mu^-\mu^+\mu^-)$ (right). 

From the $3\times1$ topology approach used as validation of the inclusive reconstruction and BDT selection, we obtain an upper limit of $2.0\times 10^{-8}$ at 90\% confidence level compatible with the Belle result. 
Table~\ref{tab:lfv} shows the Belle~II result compared with previous results from other experiments.

\begin{table}[!ht]
    \centering
    \caption{Upper limits at 90\% confidence level on the branching ratio $\mathcal{B}(\tau^- \to \mu^-\mu^+\mu^-)$ obtained in previous experiments and at Belle~II, with the corresponding integrated luminosity used for the analyses.\\}
    \begin{tabular}{l|c|c}
         \hline
         & UL at 90\% CL on $B(\tau \to 3\mu)$ &$\mathcal{L}_{int}$ [fb$^{-1}$]\\
         \hline
         Belle~\cite{lfvbelle} &$2.1 \times 10^{-8}$ &782\\
         \babar~\cite{lfvbabar} &$3.3 \times 10^{-8}$ &468\\
         CMS~\cite{lfvcms} &$2.9 \times 10^{-8}$ &131\\
         LHCb~\cite{lfvlhcb} &$4.6 \times 10^{-8}$  &2.0\\
         Belle~II &$1.9 \times 10^{-8}$ &424\\
         \hline
    \end{tabular}
    \label{tab:lfv}
\end{table}

\section{Measurement of the \texorpdfstring{$e^+e^- \to \pi^+\pi^-\pi^0$}{} cross section}
The tension between experimental results and theoretical predictions of the anomalous magnetic moment of the muon, $a_{\mu} = (g_{\mu} - 2)/2$, is generating significant interest in the community because it might arise from effects of new particles contributing to $a_{\mu}$ through virtual loops. 
The largest uncertainty in the prediction of $a_{\mu}$ is from the Hadronic Vacuum Polarization (HVP) contribution, $a^{HVP}_{\mu}$, that accounts for the 82\% of the uncertainty.

In the context of the Muon $(g-2)$ Theory Initiative~\cite{ti}, the discrepancy between theoretical calculations and experimental results exceeds 5$\sigma$, including the latest result from the Muon $(g-2)$ collaboration~\cite{g2collab}. The $5\sigma$ discrepancy reduces to $1\sigma$ when considering the predictions of $a^{HVP}_{\mu}$ based on lattice QCD~\cite{bmw} from the BMW Collaboration, and the latest result on the $e^+e^-\to\pi^+\pi^-$ cross section measurement from the CMD-3 Collaboration~\cite{cmd3}. 

The $a^{HVP}_{\mu}$ depends on the hadronic ratio $R(s) = \sigma(e^+e^- \to hadrons)/\sigma(e^+e^-\to\mu^+\mu^-)$, where $s$ is the energy of the centre-of-mass squared. Belle~II contributes to improve theoretical predictions by providing the cross section measurements for $e^+e^-\to hadrons$. For $\sqrt{s} < 1$~GeV, the two largest contributions to $a^{HVP}_{\mu}$ and its uncertainty are from the $e^+e^- \to \pi^+\pi^-$ and $e^+e^- \to \pi^+\pi^-\pi^0$ processes, which respectively contribute for the 73\% and 7\% to $a^{HVP}_{\mu}$, and for the 63\% and 15\% to the $a^{HVP}_{\mu}$ uncertainty. 

We present a measurement of the $e^+e^- \to \pi^+\pi^-\pi^0$ cross section in the energy range 0.62--3.5~GeV using an initial-state radiation (ISR) technique, performed on a subset of the Belle~II Run1 data sample, corresponding to 191~fb$^{-1}$ collected at the $\Upsilon (4S)$ resonance. The ISR technique allows to explore a wide energy range, and it is complementary to experiments that perform beam-energy scanning. The signal process is $e^+e^- \to \pi^+\pi^-\pi^0 (\gamma)$, where the $\pi^0$ is reconstructed through its $\pi^0 \to \gamma\gamma$  decay; thus signal candidates are reconstructed from two oppositely charged particles and three photons.  
The selection of signal events is mainly based on a kinematic fit to the four-momentum of the $\pi^+\pi^-\gamma\gamma\gamma$ final state, whose sum is constrained to match the four-momentum of the $e^+e^-$ beams. The $\pi^0$ signal yield in bins of $3\pi$ mass spectrum, $M_{3\pi}$, is extracted through a fit to the two-photon invariant mass, $M_{\gamma\gamma}$; thus the $\pi^0$ mass constraint is not imposed in the kinematic fit. The signal-only $M_{3\pi}$ resulting from the signal extraction fits is corrected to account for the migration of the events across different bins due to detector resolution and final-state-radiation effects.

The main backgrounds are from $e^+e^- \to \pi^+\pi^-\pi^0\pi^0(\gamma)$, $e^+e^- \to K^+K^-\pi^0(\gamma)$ and non-ISR $e^+e^- \to q\bar{q}$, whose contribution is estimated from simulation, and corrected by using data control samples. From the data-simulation comparison, we estimate a $M_{3\pi}$ bin-by-bin correction to simulation. The signal efficiency is also estimated from simulation and corrected by using data control sample. One of the main factors to signal efficiency is the $\pi^0$ reconstruction efficiency, which is obtained by reconstructing the exclusive $e^+e^- \to \omega\gamma \to \pi^+\pi^-\pi^0\gamma$ process. The $\pi^0$ efficiency is estimated both in simulation and data through the ratio $\varepsilon(\pi^0) = N_{full}/N_{partial}$, where $N_{full}$ and $N_{partial}$ are the number of fully and partially reconstructed events, respectively. In the partial reconstructed events, we only reconstruct the $\pi^+\pi^-\gamma$ system, and the $\pi^0$ is reconstructed as recoil to the $\pi^+\pi^-\gamma$ system. In this case, the recoil mass is constrained to the $\pi^0$ mass through a kinematic fit. $N_{partial}$ is estimated from a fit to the invariant mass of the system $\pi^+\pi^-\pi^{0}_{recoil}$, while $N_{full}$ is estimated from a fit to $M_{\gamma\gamma}$ in the fully reconstructed events. From data-simulation ratio, we obtain a relative correction to the signal efficiency from the $\pi^0$ efficiency of $-1.4 \pm 1.0\%$. The uncertainty on the correction factor is dominated by the uncertainty in the background contamination in data, and it is assumed as systematic uncertainty. 

\begin{figure}[!ht] 
  \centering   
  \begin{minipage}{13pc}
      \includegraphics[width=0.99\linewidth]{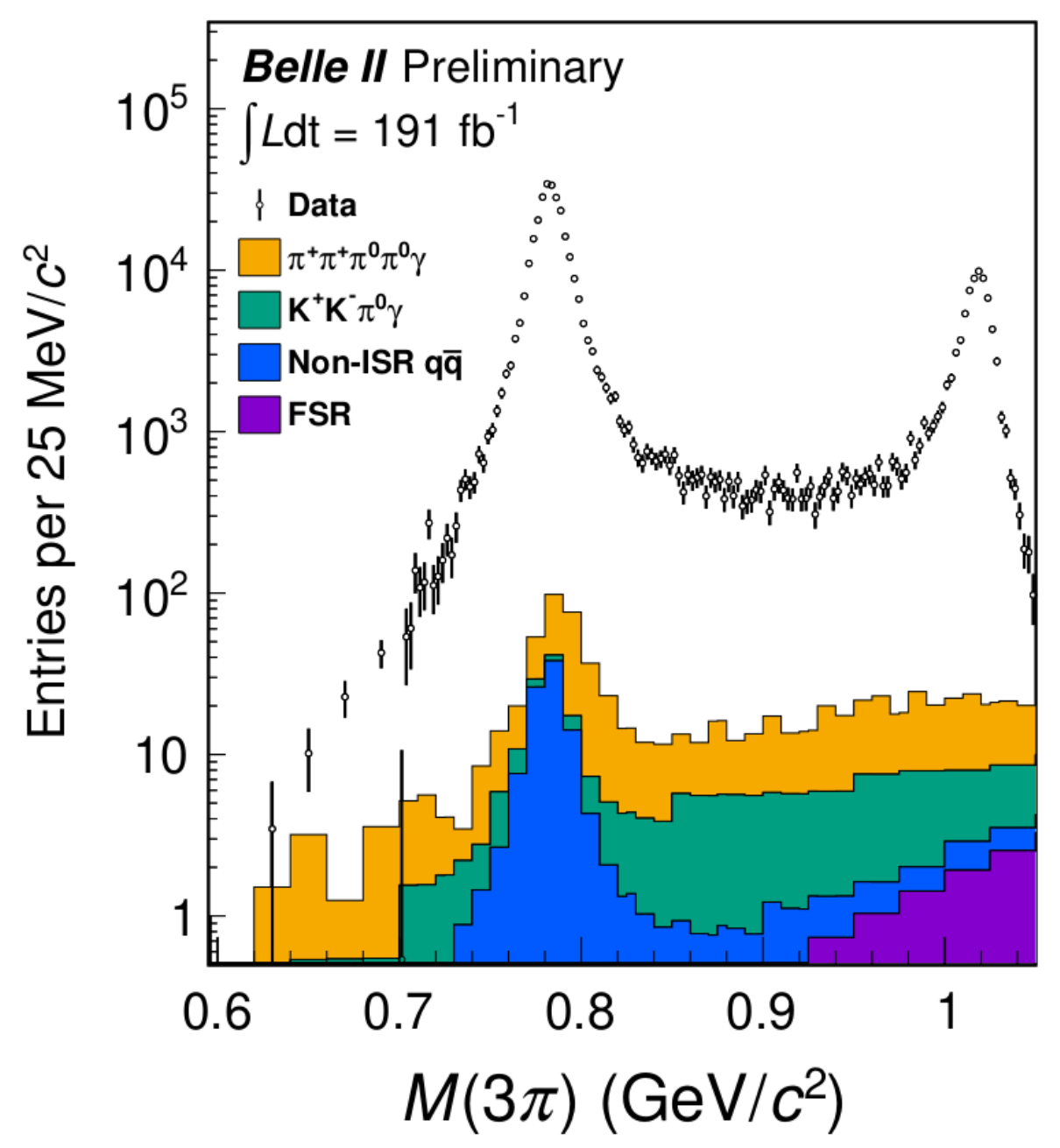}
  \end{minipage}
  \begin{minipage}{24pc}
      \includegraphics[width=0.99\linewidth]{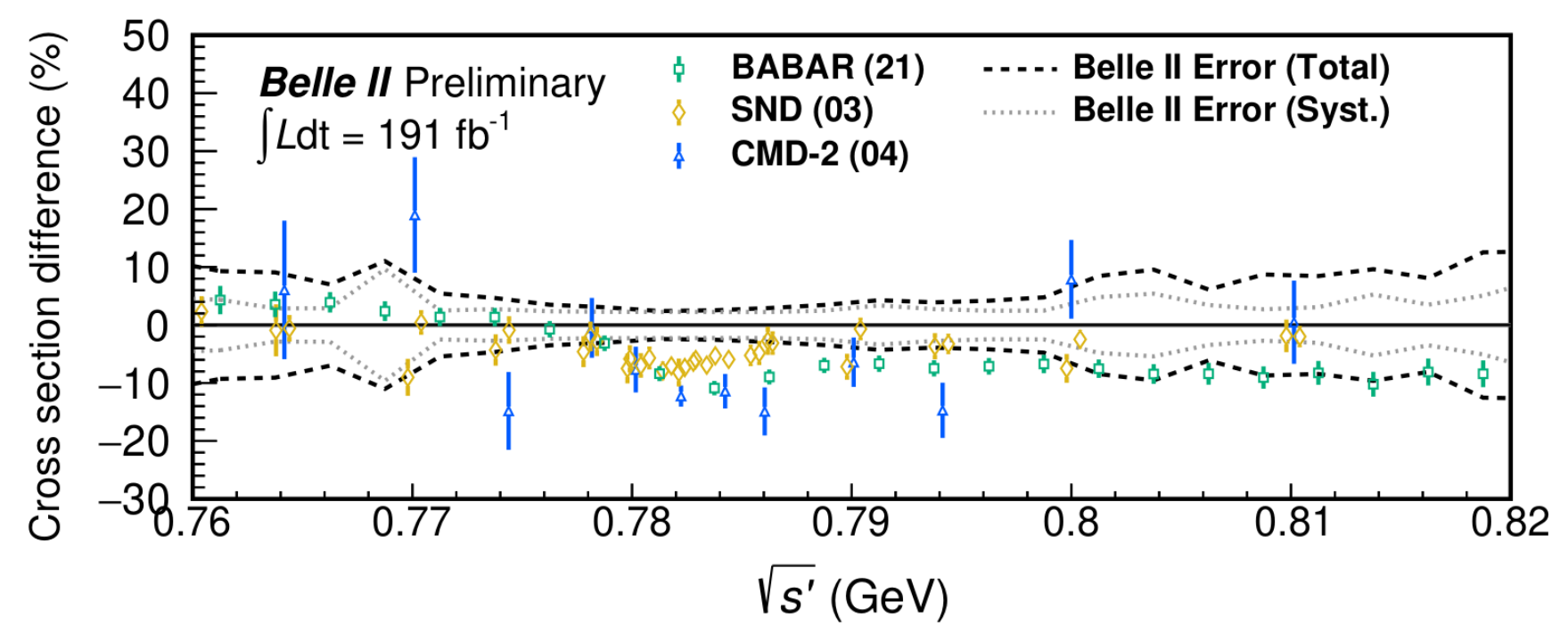} 
  \end{minipage}
  \caption{\label{fig:3pi} Left: the $M_{3\pi}$ spectrum below 1.05~GeV/$c^2$, where the $\omega$ and $\phi$ resonances are visible. The points are data, while the colored stacked histograms are the background components from simulation. Right: the difference of the cross section measured at Belle~II with respect to other experiments around the $\omega$ region.}
\end{figure}

The total systematic uncertainty is 2.2\% at the $\omega$ and $\phi$ resonances, where the cross section is larger.
It is dominated by the systematic uncertainty of 1\% on the $\pi^0$ efficiency determination, and by the systematic uncertainty of 1.2\% associated to the absence of higher-order radiative effects in the generator. At other energies, the precision on the cross section measurement is limited by the statistical uncertainty. 
The signal efficiency is 8.8\%--6.6\% depending on $M_{3\pi}$, with a total correction factor of -4.6\% determined with a accuracy of 1.6\%.  
Figure~\ref{fig:3pi} shows the $M_{3\pi}$ distribution below 1.05~GeV/$c^2$, where the $\omega$ and $\phi$ resonances are visible (left), and the difference of the cross section measured at Belle~II with respect toother experiments  around the $\omega$ region (right). In the energy region 0.78--0.80~GeV, \babar\ cross section is smaller than Belle~II. Below 0.78~GeV and above 0.80~GeV, Belle~II results are compatible with \babar\ results within the uncertainties. The resulting contribution to the muon anomalous magnetic moment, at leading order in HVP, is $a_{\mu}^{3\pi} = (48.91 \pm 0.23(stat.) \pm 1.07(syst.)) \times 10^{-10}$ in the 0.62--1.8~GeV energy range. It is 6.5\% higher than the global fits~\cite{amufit} with a 2.5$\sigma$ significance.

\section{Conclusions}
We present the latest Belle~II world-leading results from $\tau$-physics, a measurement of the $e-\mu$ lepton-flavour universality and the search for the lepton-flavour violation decay $\tau^- \to \mu^-\mu^+\mu^-$, and the measurement of the $e^+e^- \to \pi^+\pi^-\pi^0$ cross section from 0.62--3.5~GeV using an initial-state radiation technique. The measurements are performed using 362~fb$^{-1}$, 424~fb$^{-1}$, and 191~fb$^{-1}$, respectively, collected at Belle~II in Run1 (2019--2022). 

Belle and Belle~II are leading the $\tau$ searches, and are providing fundamental results from low-multiplicity event analyses, both in terms of precision measurements of the SM parameters and searches beyond the SM physics. Many frontiers of improvements are foreseen: increases in data sample size from Run2 at Belle~II, which started on February 20$^{\rm th}$ 2024, innovative analysis techniques to increase signal efficiency while keeping background under control, and reduced systematic uncertainties.


\section*{References}

\begin{thebibliography}{}

\bibitem{B2TIP}
E. Kou et al. (Belle~II Collaboration), \href{https://doi.org/10.1093/ptep/ptz106}{{\em Prog. Theor. Exp. Phys.} 2019, 123C01 (2019)}.

\bibitem{SKEKB}
K. Akai, K. Furukawa, and H. Koiso (SuperKEKB Accelerator Team), \href{https://doi.org/10.1016/j.nima.2018.08.017}{{\em Nucl. Instrum. Methods} A 907, 188 (2018)}.

\bibitem{B2TDR}
T. Abe et al. (Belle~II Collaboration), \href{https://doi.org/10.48550/arXiv.1011.0352}{arXiv:1011.0352 [physics.ins-det] (2010)}.

\bibitem{BTDR}
A. Abashian at al. (Belle Collaboration), \href{https://doi.org/10.1016/S0168-9002(01)02013-7}{{\em Nucl. Instrum. Methods} A 479, 117 (2002)}.

\bibitem{KEKB}
T. Abe et al., \href{https://doi.org/10.1093/ptep/pts102}{{\em Prog. Theor. Exp. Phys.} 2013, 03A001 (2013)},

\bibitem{taumass}
I. Adachi et al. (Belle~II Collaboration) \href{https://link.aps.org/doi/10.1103/PhysRevD.108.032006}{{\em Phys. Rev.} D 108, 032006 (2023)}

\bibitem{taula}
I. Adachi et al. (Belle~II Collaboration) \href{https://link.aps.org/doi/10.1103/PhysRevLett.130.181803}{{\em Phys. Rev. Lett.} 130, 181803 (2023)}

\bibitem{snow}
L. Aggarwal et al. (on behalf of the Belle~II US group)
\href{https://arxiv.org/abs/2207.06307}{arXiv:2207.06307 [hep-ex] (2022)}

\bibitem{lfu}
Yung-Su Tsai, \href{https://link.aps.org/doi/10.1103/PhysRevD.4.2821}{{\em Phys. Rev.} D 4, 2821 (1971)}; Erratum \href{https://link.aps.org/doi/10.1103/PhysRevD.13.771}{{\em Phys. Rev.} D 13, 771 (1976)}

\bibitem{lfucleo}
A. Anastassov et al. (CLEO Collaboration)
\href{https://link.aps.org/doi/10.1103/PhysRevD.55.2559}{{\em Phys. Rev.} D 55, 2559 (1997)}; Erratum \href{https://link.aps.org/doi/10.1103/PhysRevD.58.119903}{{\em Phys. Rev.} D 58, 119903 (1998)}

\bibitem{lfubabar}
B. Aubert et al. (\babar\ Collaboration)
\href{https://link.aps.org/doi/10.1103/PhysRevLett.105.051602}{{\em Phys. Rev. Lett.} 105, 051602 (2010)}

\bibitem{hflav}
Y. Amhis et al. (Heavy Flavor Averaging Group Collaboration), \href{https://link.aps.org/doi/10.1103/PhysRevD.107.052008}{{\em Phys. Rev.} D 107, 052008 (2023)}

\bibitem{lfvbelle}
K. Hayasaka et al. (Belle Collaboration), \href{https://doi.org/10.1016/j.physletb.2010.03.037}{{\em Phys. Lett.} B 687, 139 (2010)}

\bibitem{punzi}
G. Punzi, \href{https://doi.org/10.48550/arXiv.physics/0308063}{arXiv:physics/0308063 (2003)}

\bibitem{lfvbabar}
J. P. Lees et al. (\babar\ Collaboration), \href{https://doi.org/10.1103/PhysRevD.81.111101}{{\em Phys. Rev.} D 81, 111101 (2010)} 

\bibitem{lfvcms}
A. M. Sirunyan et al. (CMS Collaboration), \href{https://doi.org/10.1007/JHEP01(2021)163}{{\em JHEP} 01, 163 (2021)}

\bibitem{lfvlhcb}
R. Aaij et al. (LHCb Collaboration), \href{https://doi.org/10.1007/JHEP02(2015)121}{{\em JHEP} 02, 121 (2015)} 

\bibitem{ti}
T. Aoyama et al. (Muon $(g-2)$ Theory Initiative), \href{https://doi.org/10.1016/j.physrep.2020.07.006}{{\em Phys. Rep.} 887, 1 (2020)}

\bibitem{g2collab}
G. Venanzoni et al. (Muon $g-2$ Collaboration), \href{https://doi.org/10.48550/arXiv.2311.08282}{arXiv:2311.08282 [hep-ex] (2023)}

\bibitem{bmw}
S. Borsanyi et al. (BMW Collaboration), \href{https://doi.org/10.1038/s41586-021-03418-1}{{\em Nature} 593, 51–55 (2021)}.

\bibitem{cmd3}
F. V. Ignatov et al. (CMD-3 Collaboration), \href{https://doi.org/10.48550/arXiv.2302.08834}{arXiv:2302.08834 [hep-ex] (2023)}

\bibitem{amufit}
M. Hoferichter, B.-L. Hoid, B. Kubis, and D. Schuh, \href{https://link.springer.com/article/10.1007/JHEP08(2023)208}{JHEP 08, 208 (2023)}

\end{thebibliography}
\end{document}